\documentclass[prl,twocolumn,superscriptaddress,showpacs]{revtex4}
\usepackage{amsmath,amssymb,graphicx}

\begin{document}
\title{Polarization multistability of cavity polaritons}

\author{N. A. Gippius}
 \affiliation{LASMEA, CNRS, Universit\'{e} Blaise Pascal, 24
av.~des Landais, 63177 Aubi\`ere, France}
\affiliation{General Physics Institute, RAS, 38, Vavilov Street, Moscow, 119991, Russia}

\author{I. A. Shelykh}
\affiliation{International Center for Condensed Matter Physics, Universidade de Brasilia, 70904-970 Brasilia DF, Brazil}

\author{D. D. Solnyshkov}
\affiliation{LASMEA, CNRS, Universit\'{e} Blaise Pascal, 24
av.~des Landais, 63177 Aubi\`ere, France}

\author{S. S. Gavrilov}
\affiliation{LASMEA, CNRS, Universit\'{e} Blaise Pascal, 24
av.~des Landais, 63177 Aubi\`ere, France}
\affiliation{Institute of Solid State Physics, RAS, Chernogolovka, Moscow District 142432, Russia}

\author{Yuri G. Rubo}
\affiliation{Centro de Investigaci\'on en Energ\'{\i}a, Universidad
Nacional Aut\'onoma de M\'exico, Temixco, Morelos 62580, Mexico}

\author{A. V. Kavokin}
 \affiliation{School of Physics \& Astronomy, University of
Southampton, SO17 1BJ, Southampton, UK}

\author{S. G. Tikhodeev}
\affiliation{General Physics Institute, RAS, 38, Vavilov Street, Moscow, 119991, Russia}

\author{G. Malpuech}
 \affiliation{LASMEA, CNRS, Universit\'{e} Blaise Pascal, 24
av.~des Landais, 63177 Aubi\`ere, France}

\date{December 22, 2006}

\begin{abstract}
New effects of polarization multistability and polarization
hysteresis in a coherently driven polariton condensate in a
semiconductor microcavity are predicted and theoretically analyzed.
The multistability arises due to polarization-dependent
polariton-polariton interactions and can be revealed in polarization
resolved photoluminescence experiments. The pumping power required
to observe this effect is of 4 orders of magnitude lower than the
characteristic pumping power in conventional bistable optical
systems.
\end{abstract}


\maketitle


\emph{Introduction.}---Cavity polaritons are elementary excitations
of semiconductor microcavities with extremely small effective mass
$m^*$ ranging from 10$^{-4}$ to 10$^{-5}$ of the free electron mass
\cite{KavokinBook}.

In
the low density limit they behave as weakly interacting bosons and
their Bose-Einstein condensation (BEC) has been recently
claimed
\cite{Kasprzak2006}. Under resonant excitation two main nonlinear
mechanisms have been identified, the polariton parametric
scattering~\cite{Savvidis2000,Tartakovski2000,Stevenson2000,Ciuti2000,Ciuti2001,Savvidis2002}
and bistability of the polariton system \cite{Tredicucci,
Baas04,Gippius04e,Gippius04j,Whittaker2005}.
These two nonlinear
mechanisms often coexist, being of the same
origin~\cite{Gippius04e,Gippius04j}.

Another important peculiarity of the cavity
polaritons is related to their polarization or pseudospin degree of
freedom. Polaritons have two possible spin projections on the
structure growth axis, $\pm1$, corresponding to the right and left
circular polarizations of the counterpart photons.
In case of non-zero in-plane wave-vector ($k \neq 0$) these two
components are mixed by TE-TM splitting~\cite{Panzarini1999}, but we
will be interested in $k=0$ case. The mixing of the $k=0$ states
appears due to the polariton-polariton interaction, which
depends on the spin orientation. Namely, the interaction of
polaritons in triplet configuration (parallel spin projections on
the structure growth axis) is different from that of polaritons in
singlet configuration (antiparallel spin projections). The
spin-dependent polariton coupling strongly affects the predicted
superfluid properties of the polariton system
\cite{ShelykhPRL,Rubo2006} and leads to remarkable nonlinear effects
in polariton spin relaxation, such as self-induced Larmor precession
and inversion of the linear polarization during the scattering act
\cite{ShelykhPSSb}.

In this Letter we show that the interplay between the nonlinearity
caused by the polariton-polariton interactions and the polarization
dependence of these interactions results
in a remarkable \emph{multistability} of a driven polariton
condensate, contrary to the usual optical \emph{bistability} in the
spin-less nonlinear case. We consider the ground state ($k=0$)
polariton mode having a finite life-time which is coherently pumped
at normal incidence to the microcavity plane by a \emph{cw} laser of
variable intensity, frequency, and polarization. We calculate the
resulting intensity and polarization of the polariton field and show
that, for a given polarization of the pump and  depending on the
history of the pumping process, the polariton polarization can, in
general, take three different values. For instance, a linearly
polarized laser can result in a strongly right-circularly, strongly
left-circularly, or in a linearly polarized polariton state. The
whole system is therefore not simply bistable but tri- or even
multistable. It is worth noting that this conversion from linear to
circular polarization arises without any TE-TM splitting
\cite{ShelykhPRB04} and represents a new nonlinear effect.

\begin{figure}
\includegraphics[width=0.9\linewidth]{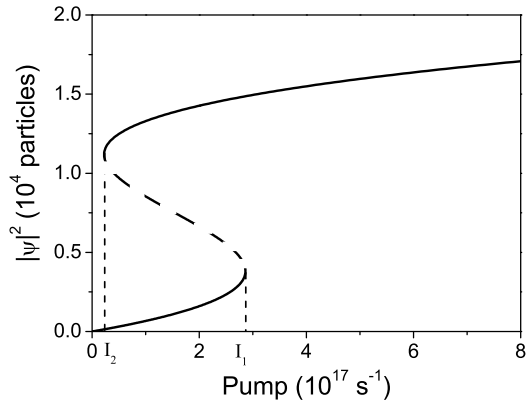} 
\caption{\label{sshape} The condensate population
$\left|\psi_\sigma\right|^2$ versus pumping intensity $|p_\sigma|^2$
for the circularly polarized pump ($\sigma=\pm1$). Dashed line shows
the unstable region. $I_1$ and $I_2$ mark the intensities of two
jumps corresponding to the increase and decrease of the
pump, respectively. The numbers shown (scaled to the excitation area
$S \approx 100~\mu\mathrm{m}^2$) are typical for cavity polariton
experiments.
}
\end{figure}

Qualitatively, the multistability can be understood as follows.
Assume that the cavity is illuminated by a laser light at normal
incidence at the frequency above the bottom of the lowest polariton
branch (LPB). At low pumping, the pump is not in resonance with the
polariton eigenstate, so that the population of the driven mode
remains low. At higher pumping, polariton-polariton interactions
lead to the blue-shift of the LPB, so that it approaches the pump
laser frequency. At resonance, the population jumps up abruptly. If
the pumping power is then decreased, the population of the polariton
mode jumps down back, but at a lower threshold. As a result the
typical S-shape dependence of the intra-cavity field on the pumping
intensity appears as Fig.\ \ref{sshape} shows, which means the
formation of a hysteresis cycle. The additional polarization degree
of freedom makes this picture much more complex and rich. At
elliptically polarized pumping the hysteresises for right- and
left-circularly polarized components of the field do not coincide,
which results in a sequence of polarization jumps and
multistability.

\emph{Coherent polarization evolution.}---We consider a
semiconductor
microcavity in the strong coupling regime pumped by a laser light perpendicular
to the cavity plane ($k=0$), with spatially uniform
and slowly changing with time intensity
$p_\sigma$ and frequency $\omega$ near the bottom of the LPB, where $\sigma=\pm 1$ is the circular polarization
index. We neglect the non-parabolicity of the polariton dispersion, and the wave vector dependence of exciton
and photon fractions. These approximations are well justified since
the effects we discuss are taking place in a narrow energy and wave
vector range close to the polariton ground state at $k=0$ with the
eigenfrequency $\omega_0$.

The $k=0$ polariton wave function $\psi_\sigma$ satisfies
the driven spinor Gross-Pitaevskii equation \cite{ShelykhPRL,Carusotto2004},
which in the quasi-stationary case can be written as (we set
$\hbar=1$)
\begin{equation}
\label{GPuniform}
 \left[\omega_0-\omega-\frac{\mathrm{i}}{\tau}
 + \alpha_1|\psi_\sigma|^2 + \alpha_2|\psi_{-\sigma}|^2\right]\psi_\sigma
 + \frac{p_\sigma}{\sqrt{4\tau}}=0,
\end{equation}
where $\tau$ is the polariton lifetime, $\alpha_{1(2)}$ is the
matrix element of polariton-polariton interaction in the
triplet(singlet) configuration, respectively.
The exchange
interaction is suppressed for polaritons in the singlet
configuration since it involves energetically split-off dark
excitons as intermediate states. As a result, there is an
attraction between polaritons of opposite sign, $\alpha_2<0$,
and polariton-polariton
interaction is strongly anisotropic  $|\alpha_2|\ll
\alpha_1$ \cite{Renucci2005}.

At thermodynamic equilibrium the polariton BEC leads to the
formation of the linearly polarized condensate, which minimizes the
free energy of the system \cite{ShelykhPRL}. In the case of
\textit{cw} driven system the situation is qualitatively different.
The polariton system is intrinsically out of equilibrium and its
state is now driven by the pump. The parameters
of the driven mode, however, are not uniquely defined. E.g., with a
linearly polarized pump the polariton condensate can change its
polarization either to the left- or to the right-circular, increasing
this way the blue-shift and entering into resonance with the pump.
The left and right circularly polarized condensates can appear in
this case with equal probabilities.

Equation~(\ref{GPuniform}) can be used to relate the occupation of the
condensate $N=|\psi_{+1}|^2+|\psi_{-1}|^2$ and the degree of its
circular polarization $\rho_c=(|\psi_{+1}|^2-|\psi_{-1}|^2)/N$ with
the pump intensity $I=|p_{+1}|^2+|p_{-1}|^2$ and the pump
polarization degree $\rho_p=(|p_{+1}|^2-|p_{-1}|^2)/I$. Namely,
\begin{multline}
\label{IvsNa}
 I/4\tau=[\Omega^2+\tau^{-2}
    +(\alpha_1-\alpha_2)(1-\rho_c^2)\Omega N \\
    +(1/4)(\alpha_1-\alpha_2)^2(1-\rho_c^2)N^2]N,
\end{multline}
\begin{equation}
\label{IvsNb}
 \rho_pI/4\tau=[\Omega^2+\tau^{-2}
 - (1/4)(\alpha_1-\alpha_2)^2(1-\rho_c^2)N^2]\rho_cN,
\end{equation}
where $\Omega=\omega-\omega_0-\alpha_1N$.

It follows from Eqs.\ (\ref{IvsNa}) and (\ref{IvsNb}) that only for
the pump with a pure circular polarization, $\rho_p=\pm 1$, the
polarizations of the polariton system and the pump coincide. In this
case the present spinor model can be reduced to the scalar model
\cite{Gippius04e,Gippius04j}. For elliptical pumping the
polarizations of the polariton system and of the laser are different
due to the different blue shifts for right and left circular
polarized components. Due to the effect of the self-induced Larmor
precession \cite{ShelykhPSSb} the polarization ellipse of the driven
mode is rotated with respect to the polarization ellipse of the pump
by some angle which depends on the pump intensity. Moreover, in this
regime even the signs of circular polarization degrees of the
polaritons and the pump can be different, as we show in the next
subsection.

\emph{Multistability and hysteresis.}---The polarization and
occupation of the driven mode, $\rho_c$ and $N$, can be determined
self-consistently from Eqs.\ (\ref{IvsNa}) and (\ref{IvsNb}), which
can have several solutions depending on the value and polarization of the pump.

The condensate population $|\psi_{\pm 1}|^2$
as a function the pump intensity $|p_{\pm 1}|^2$ for the case of fully
circular polarized excitation (when the polarization of the driven
mode coincides with that of the pump) is shown in Fig.\
\ref{sshape}. The energy of the laser $\omega$ is chosen above the
energy of the bare polariton state, $\omega-\omega_0=3$ meV, so that
the curve shows the classical S-shape. The matrix element $\alpha_1
= 6 x E_b a_\mathrm{B}^2/S $, where $a_\mathrm{B}=100\;\AA$ is the
two dimensional exciton Bohr radius, $E_b=8$ meV is the exciton
binding energy, $x=1/4$ is the squared exciton fraction, and
$S=100~\mu\mathrm{m}^2$ is the laser spot area. The polariton
life-time is $\tau =2$ ps. These parameters are typical for a GaAlAs
microcavity. We have denoted by $I_1$ the laser intensity
corresponding to the bending point of the S-shape taking place with
the increase of intensity, and $I_2$ is the laser intensity
corresponding to the bending point in the case of the intensity
decrease.

\begin{figure}
\includegraphics[width=1.0\linewidth]{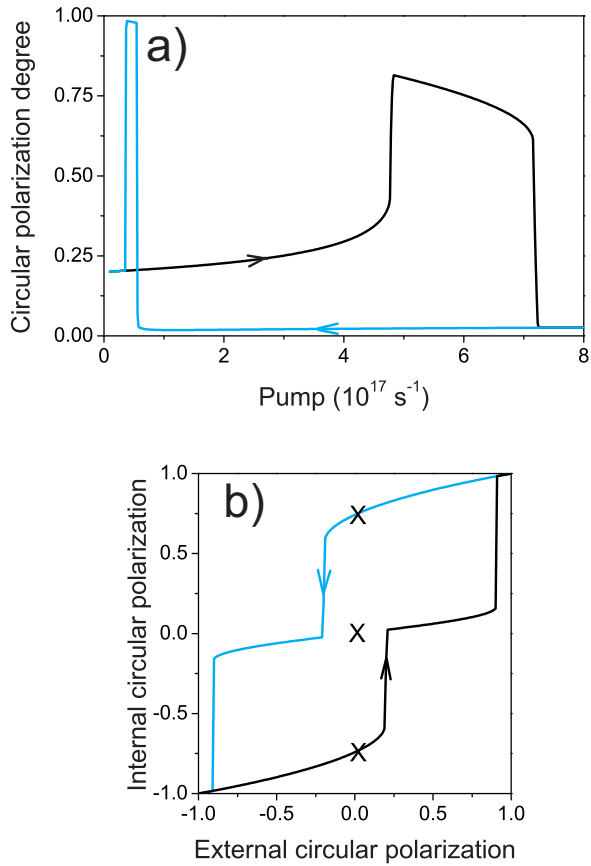}
\caption{\label{fig2} (a) Circular polarization degree of the
condensate versus external pumping intensity for slightly elliptical
pump ($\rho_p=0.2$). Arrows show the direction in which the pump
intensity is changed. (b) Circular polarization degree of the driven
mode $\rho_c$ versus circular polarization degree of the pump
$\rho_p$. Arrows show the direction in which the pump polarization
degree is changed. The pump intensity is just above $I_1$. Crosses
mark the stability points, which correspond to $\rho_p=0$.}
\end{figure}

The evolution of the condensate polarization can be conveniently
illustrated and understood considering the case $\alpha_2=0$, when
Eqs.\ (\ref{GPuniform}) for two values of $\sigma$ are simply
decoupled and the two circular components evolve independently
\cite{alphanote}. The change of the $\rho_c$ with the total pump
intensity is shown in Fig.\ \ref{fig2}(a) for the elliptically
polarized pump with $\rho_p=0.2$, so that $\sigma^{+}$ component slightly exceeds
$\sigma^{-}$ one.
As one increases the intensity of the pump, the solution moves along
the lower branch of the S-curve for both $\sigma^{+}$ and
$\sigma^{-}$ components. However, the $\sigma^{+}$ component, since
it dominates, reaches the threshold intensity $I_1$ first. The
corresponding intensity of the polariton field jumps abruptly to the
upper branch. At the same time, the intensity of the $\sigma^{-}$
pump has not yet reached the $I_1$ bending point. So, the jump of
the total polariton density is accompanied by a jump of the circular
polarization degree. If we increase further the intensity of the
laser, the intensity of the $\sigma^{-}$ mode also reaches $I_1$.
The polariton population increases again, but this now results in an
abrupt decrease of the circular polarization degree of the driven
mode. If we now reduce the intensity, the reversed process takes
place at the pumping intensity $I_2$ so that hysteresises in both
the occupation and polarization power dependencies appear.

Fig.\ \ref{fig2}(b) shows another interesting configuration, where
the laser intensity $I$ is kept constant in the domain $I>I_1$ and
$I_1>I/2>I_2$. The cyan line shows the change of the circular
polarization degree of the driven mode $\rho_c$ induced by the laser
initially polarized $\sigma^{+}$ and whose polarization is
progressively rotated towards the $\sigma^{-}$ polarization. One can
observe a weak decrease of $\rho_c$, which however remains quite
high even when the pumping is linearly polarized. This is
due to the fact that the $\sigma^{+}$ component remains  on
the upper branch of the S-curve whereas the $\sigma^{-}$ drops on
the lower branch. Then there are two jumps of polarization
corresponding to the jumps of $\sigma^{+}$ and $\sigma^{-}$
components of the polariton population, and finally the polarization becomes fully
$\sigma^{-}$. The black line shows the evolution of $\rho_c$ with
the inverse change of the pump polarization. The stable points
corresponding to full linear polarization of the laser are marked
with crosses in this Figure. One can see that the internal
polarization can be either nearly $\sigma^{+}$, either nearly
$\sigma^{-}$, or fully linear. The latter case is in fact
degenerate. There can be two stable driven mode occupations $N$ for
the same value of the external laser intensity $I$.

\begin{figure}
\includegraphics[width=0.9\linewidth]{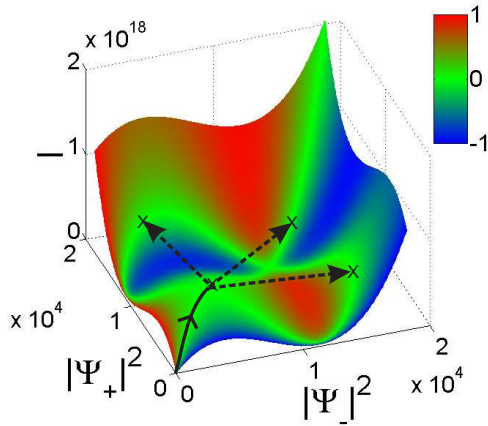}
\caption{ \label{coolfig} The pump intensity $I$ and polarisation (color) versus the
circular-polarized components of the driven mode $|\psi_{\pm 1}|^2$.
Color shows the polarization of the pump $\rho_p$ (bright green
corresponds to linear polarization). The crosses mark the four
stable points for the driven mode corresponding to the same linearly
polarized pump intensity. Arrows show the three possible jumps in
case if the pump intensity is slightly increased.}
\end{figure}

Fig.\ \ref{coolfig} shows the functional dependence between $\sigma^{+}$ and $\sigma^{-}$
components of the polariton population and the intensity and polarisation of pump calculated
accounting for the coupling between polaritons with opposite spins described by $\alpha_2=-0.1\alpha_1$.
This value of $\alpha_2$ corresponds to recent estimations of Ref.\
\cite{Renucci2005}. The circular polarization degree of the pump
laser is represented by the color of the surface of solution and the
intensity of the pump is on the vertical $z$-axis. The linear part
of the polarization of the laser is kept aligned along x-direction.

The green bright areas correspond to nearly linearly polarized pumping.
If one increases the intensity of the pumping laser, keeping it
linearly polarized, the system follows the black solid line and black arrows
shown in the figure. One can see that from the critical point at the end of the solid black line the system can jump
into three possible stable points (shown by crosses). One of them
corresponds to linearly polarized state and two others to nearly
right- and left-circularly polarized states. The choice of the final state by the system is random and
is triggered by fluctuations.

Note also that the jump of the system to the circularly polarized final
state induces a red shift of the
cross-circular component because of the negative sign of $\alpha_2$.
The red shift drives this component out of resonance with the pump, which leads to stronger
polarisation of the final state. The positive feedback in polarisation of
the condensate would not take place for the case $\alpha_2=0$ where only a linearly
polarized driven mode would be formed. Experimentally, we expect this effect to have
a key impact on the polarization measurements performed with
resonantly excited microcavities. It will result in a random sign of
the observed polariton polarization changing from one experiment to
another.

It should be noted that polarization multistability and chaos in
nonlinear optical systems have been studied for more than 30 years (see, e.g.,
\cite{Zheludev}). A number of intriguing nonlinear effects have been
predicted and observed in different systems, including four-wave
mixing in anisotropic crystals \cite{Yumoto, ZhelExp}, magnetic
cavities \cite{Jonsson} and VCSELs \cite{VcselPol}. However, in the
systems investigated previously extremely high powers (10~MW/cm$^2$
\cite{Zheludev}) are required for observation of polarization
multistability effects linked to optical nonlinearities. The
advantage of the microcavities is that in the strong coupling regime
the nonlinear threshold corresponds to much lower pumping powers
(e.g. 650~W/cm$^2$ \cite{Kulakovskii}) which really opens the way to
practical realization of new applications like data encryption and
communication \cite{Fischer}.

\emph{Conclusions.}---We have analyzed the polarization of the
spinor polariton system pumped by a \textit{cw} laser of variable
polarization at $k=0$ by solving the polarization dependent
Gross-Pitaevskii equation. We have shown that for a given
polarization of the pumping laser the polariton polarization can
take three different values. For instance at linear pumping
the resulting polariton state can be right circularly, left
circularly or linearly polarized. In realistic cases, the driven
mode polarization is random and is triggered by the fluctuations of
the external laser.

The multistability arises from the
interplay between the anisotropy of the polariton-polariton
interaction and the bistable behavior of cavity polaritons.

This work was supported by Marie-Curie RTN "Clermont2", the STREP
"STIMSCAT", the ANR Chair of Excellence Program, and RAS Program
''Strongly correlated electrons in semiconductors''. YGR acknowledges
the support from the grant IN107007 of DGAPA-UNAM. The authors thank
V.D. Kulakovskii, K.V. Kavokin and  T. Liew for useful discussions.


\begin{thebibliography}{99}
\bibitem{alphanote} The case $\alpha_2=0$ should not be mixed with
the case $\alpha_1=\alpha_2$. The polarization multistability is
present in the former, while the latter is identical to the
spin-less model, as seen from Eqs.\ (\ref{IvsNa}) and (\ref{IvsNb}).
\end{thebibliography}
\end{document}